\documentclass[aip,reprint]{revtex4-1}

\usepackage{amsmath}
\usepackage{amssymb}

\usepackage{textcomp}
\usepackage{graphicx}
\usepackage{dcolumn}
\usepackage{bm}
\usepackage{upgreek}
\graphicspath{{images/}}
\usepackage{comment}
\usepackage{siunitx}

\draft 

\begin{document}


\title{Forward volume magnetoacoustic spin wave excitation with micron-scale spatial resolution} 



\author{M. K\"u\ss{}}
 \email{matthias.kuess@physik.uni-augsburg.de.}
 \affiliation{Experimental Physics I, Institute of Physics, University of Augsburg, 86135 Augsburg, Germany\looseness=-1}
 
\author{F. Porrati}
 \affiliation{Institute of Physics, Goethe University, 60438 Frankfurt am Main, Germany}

\author{A. H{\"o}rner}
 \affiliation{Experimental Physics I, Institute of Physics, University of Augsburg, 86135 Augsburg, Germany\looseness=-1}

\author{M. Weiler}
 \affiliation{Fachbereich Physik and Landesforschungszentrum OPTIMAS, Technische Universit{\"a}t Kaiserslautern, 67663 Kaiserslautern, Germany}
 
\author{M. Albrecht}
 \affiliation{Experimental Physics IV, Institute of Physics, University of Augsburg, 86135 Augsburg, Germany\looseness=-1}
 
\author{M. Huth}
 \affiliation{Institute of Physics, Goethe University, 60438 Frankfurt am Main, Germany}
 
\author{A. Wixforth}
 \affiliation{Experimental Physics I, Institute of Physics, University of Augsburg, 86135 Augsburg, Germany\looseness=-1}



\begin{abstract}
The interaction between surface acoustic waves (SAWs) and spin waves (SWs) in a piezoelectric-magnetic thin film heterostructure yields potential for the realization of novel microwave devices and applications in magnonics.
In the present work, we characterize magnetoacoustic waves in three adjacent magnetic micro-stripes made from CoFe+Ga, CoFe, and CoFe+Pt with a single pair of tapered interdigital transducers (TIDTs).
The magnetic micro-stripes were deposited by focused electron beam-induced deposition (FEBID) and focused ion beam-induced deposition (FIBID) direct-writing techniques.
The transmission characteristics of the TIDTs are leveraged to selectively address the individual micro-stripes.
Here, the external magnetic field is continuously rotated out of the plane of the magnetic thin film and the forward volume SW geometry is probed with the external magnetic field along the film normal.
Our experimental findings are well explained by an extended phenomenological model based on a modified Landau-Lifshitz-Gilbert approach that considers SWs with nonzero wave vectors. 
Magnetoelastic excitation of forward volume SWs is possible because of the vertical shear strain $\varepsilon_{xz}$ of the Rayleigh-type SAW.
\end{abstract}

\maketitle 



%
%

%


\section{Introduction}

Over the last decade, increasing attention has been paid to the resonant coupling between surface acoustic waves (SAWs) and spin waves (SWs)~\cite{Bozhko.2020, Li.2021, Yang.2021}. 
On the one hand, magnetoacoustic interaction opens up the route toward energy-efficient SW excitation and manipulation in the field of magnonics~\cite{A.A.Serga.2010}. 
On the other hand magnetoacoustic interaction greatly affects the properties of the SAW, which in turn can be used to devise new types of microwave devices such as magnetoacoustic sensors~\cite{Chiriac.2001,A.Kittmann.2018} or microwave acoustic isolators~\cite{Kittel.1958, Lewis.1972,Sasaki.2017, A.HernandezMinguez.2020, Tateno.2020, Ku.2020, Verba.2018, R.Verba.2019}.
High flexibility in the design of these devices is possible since the properties of the SWs can be varied in a wide range of parameters.
For instance, the SW dispersion can be reprogrammed by external magnetic fields or electrical currents~\cite{R.A.Gallardo.2019, Ishibashi.2020} and more complex design of the magnet geometry~\cite{Krawczyk.2014, Jaris.2020} or use of multilayers~\cite{R.Verba.2019,Shah.2020,Ku.2021b,Matsumoto.2022} allow for multiple dispersion branches with potentially large nonreciprocal behavior.
Vice versa, SAW-SW interaction can be also used as an alternative method to characterize magnetic thin films, SWs, and SAWs~\cite{Xu.2020,Ku.2020,Ku.2021,Ku.2021b}.
Design of future magnetoacoustic devices can benefit from the fact that SAW technology is well-developed and already employed in manifold ways in our daily life~\cite{Campbell.1998, Lange.2008, Franke.2009, Morgan.2007}. 
Efficient excitation and detection of SAWs with metallic comb-shaped electrodes - so-called interdigital transducers (IDTs) - is possible on piezoelectric substrates. 
For example, acoustic delay lines with low insertion losses of about \SI{6}{dB} at \SI{4}{GHz} have been realized~\cite{Yamanouchi.1992}.
Fundamental limitations in the SAW excitation efficiency are mainly given by interaction with thermal phonons, spurious excitation of longitudinal acoustic waves in the air and non-linear effects at high input power~\cite{Morgan.2007, Williamson.1974}. 
%
So far, IDTs which excite SAWs homogeneously over the whole aperture have been used in resonant magnetoacoustic experiments. Apart from Refs.~\cite{Dreher.2012, Thevenard.2014}, these studies have been performed with an external magnetic field which was exclusively oriented in the plane of the magnetic thin film.

\begin{figure}
\includegraphics[width = 0.45\textwidth]{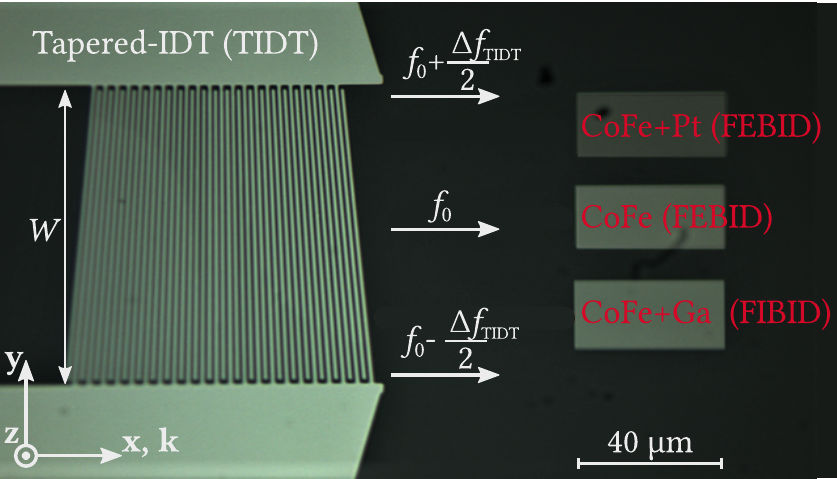}
\caption{
Optical micrograph of the fabricated device.
Rayleigh-type SAWs are excited on the piezoelectric substrate LiNbO$_3$ by a tapered-IDT (TIDT) within a wide range of frequencies $f_0-\frac{\Delta f_\text{TIDT}}{2}, \ldots, f_0+\frac{\Delta f_\text{TIDT}}{2}$.
In dependence of the applied frequency, SWs can be magnetoacoustically excited in one of the three different magnetic micro-stripes which were deposited by FEBID and FIBID.
Magnetoacoustic transmission measurements are performed by a pair of TIDTs.
\label{fig:1}
}
\end{figure}

Here, we experimentally demonstrate targeted magnetoacoustic excitation and characterization of SWs in the forward volume SW geometry with micron-scale spatial resolution. To do so, magnetoacoustic transmission measurements are performed with one pair of tapered interdigital transducers (TIDTs) at three different magnetic micro-stripes, as shown in Fig.~\ref{fig:1}.
%
This study is carried out in different geometries in which the external magnetic field is tilted out of the plane of the magnetic thin film. 
We demonstrate that magnetoelastic excitation of SWs is possible even if the static magnetization is parallel to the magnetic film normal - which is the so-called forward volume spin wave (FVSW) geometry - thanks to the vertical shear strain component $\varepsilon_{xz}$ of the Rayleigh-type SAW.
The experimental results are simulated with an extended phenomenological model, that takes the arbitrary orientation of the external magnetic field and magnetization into account.
%

The magnetic micro-stripes with lateral dimensions of about $\SI{20}{\micro m} \times \SI{40}{\micro m}$ and different magnetic properties were deposited by focused electron beam-induced deposition (FEBID) and  focused ion beam-induced deposition (FIBID). One particular advantage of using the direct-write approach~\cite{Huth.2018,Huth.2021} to fabricate the micro-stripes is the ease with which the magnetic properties can be tailored, such as the saturation magnetization~\cite{Bunyaev.2021}.
Moreover, direct-write capabilities make the fabrication of complex 3D magnetic structures on the nano-scale possible. Applications in magnonics are, for instance, 3D nanovolcanoes with tunable higher-frequency eigenmodes~\cite{Dobrovolskiy.2021}, 2D and 3D magnonic crystals with SW bandgaps~\cite{Krawczyk.2008, Gubbiotti.2019}, SW beam steering via graded refractive index, and frustrated 3D magnetic lattices~\cite{May.2019, FernandezPacheco.2020}.

\section{Theory}

\begin{figure}
\includegraphics[width = 0.48\textwidth]{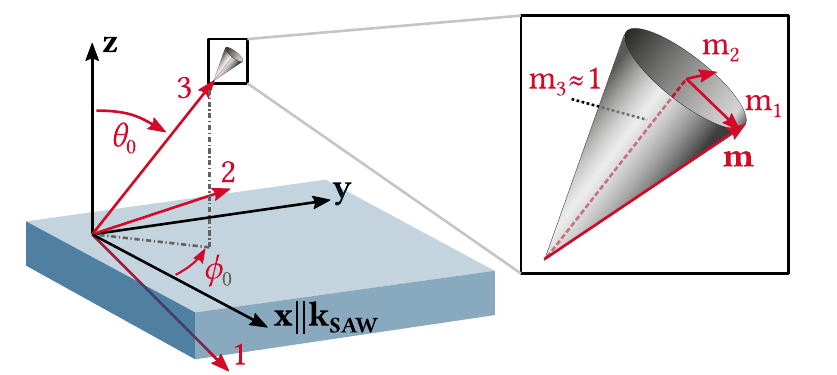}
\caption{
Relation between the coordinate systems employed.
The $(x,y,z)$ frame of reference is defined by the SAW propagation direction and the surface normal. 
We employ the (1,2,3) coordinate system to solve the LLG equation. Hereby, the 3-direction corresponds to the equilibrium magnetization orientation and the 2-direction is always aligned in the plane of the magnetic film.
The inset shows the precession cone of the magnetization, with the transverse magnetization components $m_1$ and $m_2$.
The coordinate system is taken from Ref.~\cite{Dreher.2012}.
\label{fig:2}
}
\end{figure}


A surface acoustic wave is a sound wave propagating along the surface of a solid material with evanescent displacement normal to the surface.
Density, surface boundary conditions, elastic, dielectric, and potentially piezoelectric properties of the material mainly determine if and which SAW mode can be launched. Typical SAW modes on homogeneous substrates show a linear dispersion with a constant propagation velocity of about $c_\text{SAW} = \SI{3500}{m/s}$~\cite{Morgan.2007}.
We use a standard Y-cut Z-propagation LiNbO$_3$ substrate, which gives rise to a Rayleigh-type SAW. On the substrate surface, this SAW mode causes a retrograde elliptical lattice motion in a plane defined by the SAW propagation direction and the surface normal~\cite{Rayleigh.1885, Morgan.2007}.

An optical micrograph of the fabricated magnetoacoustic device is shown in Fig.~\ref{fig:1}. Rayleigh-type SAWs can be excited in a frequency range between $f_0-\frac{\Delta f_\text{TIDT}}{2}, \ldots, f_0+\frac{\Delta f_\text{TIDT}}{2}$, which corresponds to different positions of the TIDT along the length of its aperture $W$.
To describe the magnetoacoustic transmission of the three different magnetic thin films, we extend the phenomenological model of Dreher et al.~\cite{Dreher.2012} and Küß et al.~\cite{Ku.2020} in terms of magnetoacoustically excited SWs with nonzero wave vector and arbitrary orientation of the equilibrium magnetization direction, as is detailed next.

\subsection{Magnetoacoustic driving fields and SAW transmission}

In the following, we use the $(x,y,z)$ coordinate system shown in Fig.~\ref{fig:2}~\cite{Dreher.2012}. The $x$- and $z$-axes are parallel to the wave vector $\mathbf{k}_\text{SAW} = k \hat{\mathbf{x}}$ of the SAW and normal to the plane of the magnetic micro-stripes, respectively. 
The equilibrium direction of the magnetization $\bf M$ and the orientation of the external magnetic field $\bf H$ are specified by the angles $(\theta_0, \phi_0)$ and $(\theta_H, \phi_H$).
Here, $\theta_0$ and $\phi_0$ are calculated by minimization of the static free energy. For that, we take the external magnetic field $\bf H$, thin film shape anisotropy $M_s \hat{\bf z}$ with saturation magnetization $M_s$, and a small uniaxial in-plane anisotropy $H_\text{ani}$, which encloses an angle $\phi_\text{ani}$ with the $x$-axis, into account~\cite{Dreher.2012,Ku.2020}.
Because the characterized magnetic thin films are relatively~\cite{Ku.2020} thick ($d \geq \SI{24}{nm}$), we neglect the surface anisotropy.
The SAW-SW interaction can be described by effective dynamic magnetoacoustic driving fields, which exert a torque on the static magnetization~\cite{Weiler.2011}. 
The resulting damped precession of ${\bf M}$ is then determined by the Landau--Lifshitz--Gilbert equation for small precession amplitudes. To this end, we introduce the rotated $(1,2,3)$ Cartesian coordinate system in Fig.~\ref{fig:2}. The $3$-axis is parallel to ${\bf M}$ and the $2$-axis is aligned in the film plane~\cite{Weiler.2011}.
In this phenomenological model, it is assumed that the frequencies $f$ and wave vectors $k$ of SAW and SW are identical~\cite{Gowtham.2015,Ku.2020}.
Furthermore, only magnetic films with small thicknesses $|k| d \ll 1$ and homogeneous strain in the $z$-direction of the magnetic film are considered~\cite{Dreher.2012,Ku.2020}. 

The effective magnetoacoustic driving field as a function of SAW power in the (1,2) plane can be written~\cite{Ku.2020} as
\begin{equation}
 \mathbf{h}(x,t) =
    \begin{pmatrix}
        \tilde{h}_1 \\
        \tilde{h}_2
    \end{pmatrix}
 \sqrt{ \frac{k}{R \, c_\text{SAW} W}} \sqrt{P_\text{SAW}(x)}~\text{e}^{i(kx-\omega t)}.
 \label{eq:drivingfields1}
\end{equation}
Here, $\omega = 2 \pi f$ and $c_\text{SAW}$ are the angular frequency and propagation velocity of the SAW, $W$ is the width of the aperture of the TIDT, and the constant $R=\SI{1.4E11}{J/m^3}$~\cite{Robbins.1977}. 
 The normalized effective magnetoelastic driving fields $\tilde{h}_1$ and $\tilde{h}_2$ of a Rayleigh wave with strain components $\varepsilon_{kl= xx,zz,xz} \neq 0$ are~\cite{Dreher.2012,Ku.2020}
\begin{eqnarray}
\begin{pmatrix}
    \tilde{h}_1 \\
    \tilde{h}_2
\end{pmatrix}
 =
 \frac{2}{\mu_0}
 \bigg[
    &&  b_1 \tilde{a}_{xx}
        \begin{pmatrix}
            -\sin{\theta_0} \cos{\theta_0} \cos^2{\phi_0} \\
            \sin{\theta_0} \sin{\phi_0} \cos{\phi_0}
        \end{pmatrix}
    \nonumber\\
    &&+ b_1 \tilde{a}_{zz}
        \begin{pmatrix}
            \sin{\theta_0} \cos{\theta_0} \\
            0
        \end{pmatrix}
    \nonumber\\
    &&+ b_2 \tilde{a}_{xz}
        \begin{pmatrix}
            -\cos{(2 \theta_0)} \cos{\phi_0} \\
            \cos{\theta_0} \sin{\phi_0}
        \end{pmatrix}
 \bigg],
\label{eq:drivingfields2}
\end{eqnarray}
where $b_{1,2}$ are the magnetoelastic coupling constants for cubic symmetry of the ferromagnetic layer~\cite{Dreher.2012,Kittel.1958}, $\tilde{a}_{kl} = \varepsilon_{kl,0} / (|k| |u_{z,0}|)$ are the normalized amplitudes of the strain, and $\varepsilon_{kl,0}$ are the complex amplitudes of the strain. Furthermore, $u_{z,0}$ is the amplitude of the lattice displacement in the $z$-direction. 
For the sake of simplicity, we neglect non-magnetoelastic interaction, like magneto-rotation coupling~\cite{Maekawa.1976, Xu.2020, Ku.2020}, spin-rotation coupling~\cite{Matsuo.2011, Matsuo.2013, Kobayashi.2017} or gyromagnetic coupling \cite{Kurimune.2020}.
In contrast to previous magnetoacoustic studies~\cite{Gowtham.2015,A.HernandezMinguez.2020,Xu.2020,Ku.2020,Ku.2021,Ku.2021b,Duquesne.2019} where the equilibrium magnetization direction was aligned in the plane of the magnetic film ($\theta_0=\SI{90}{\degree}$), the strain component $\varepsilon_{zz}$ results in a modified driving field for geometries with $\theta_0 \neq \SI{90}{\degree}$.

\begin{figure}
\includegraphics[width = 0.48\textwidth]{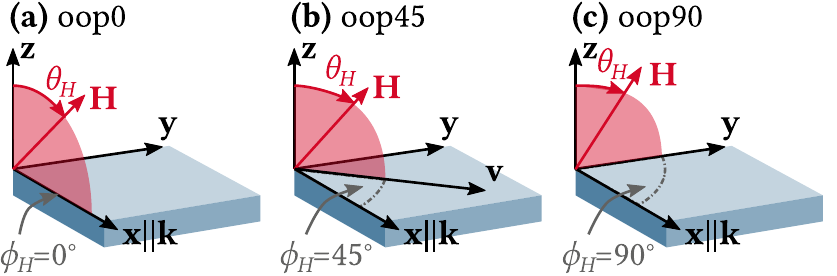}
\caption{
The magnetoacoustic transmission is studied in the three geometries oop0, oop45, and oop90, which are defined by the polar angle $\phi_H$ of the external magnetic field $\mathbf{H}$. Hereby, $\mathbf{H}$ is tilted with respect to the $z$-axis by the azimuthal angle $\theta_H$.
\label{fig:3}
}
\end{figure}

\begin{figure}
\includegraphics[width = 0.48\textwidth]{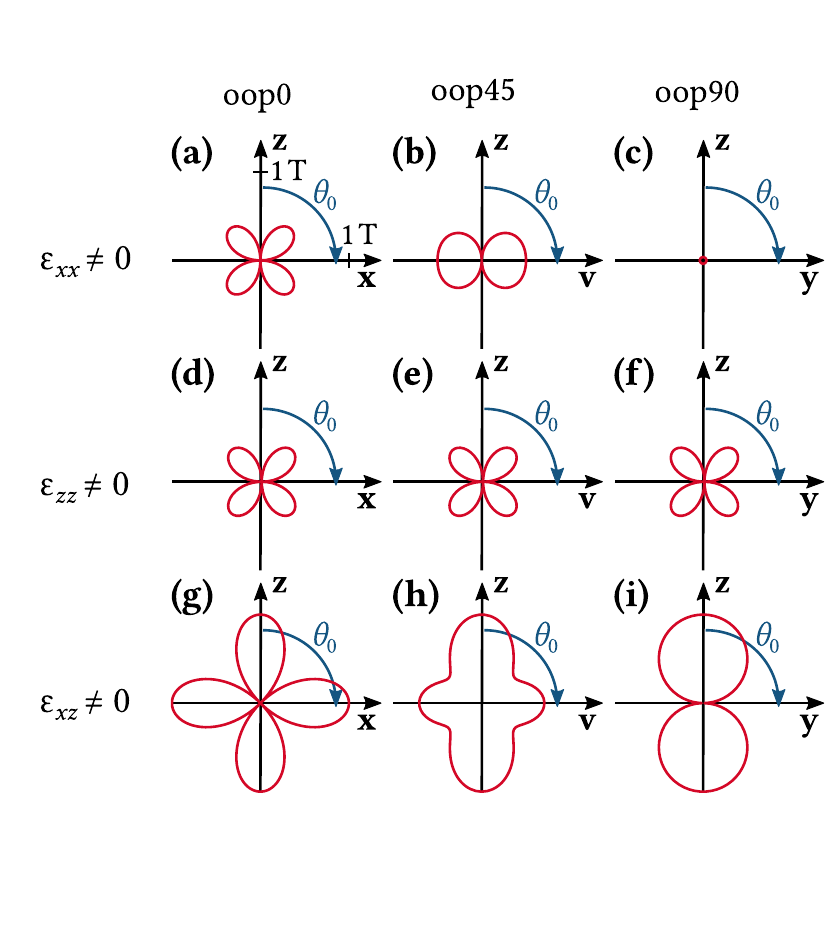}
\caption{
Polar plot of the normalized driving field’s magnitude $|\mu_0 \tilde{\mathbf{h}} (\theta_0)|$ for the relevant strain components $\varepsilon_{xx}, \varepsilon_{zz}$, and $\varepsilon_{xz}$ and for the different geometries oop0, oop45, and oop90, assuming $\phi_0 = \phi_H$. 
The distance from the origin indicates for all panels the normalized magnitude of the driving field. Thereby, the driving field was calculated by Eq.~\eqref{eq:drivingfields2} with $2 b_{1,2} \tilde{a}_{kl} = \SI{1}{T}$.
This diagram extends Fig.~4 of Ref.~\cite{Dreher.2012} by panels (c), (d), (e), (f), and (i).
\label{fig:4}
}
\end{figure}

In the experiments, we characterize SAW-SW interaction for the three geometries depicted in Fig.~\ref{fig:3}. The oop0-, oop45-, and oop90-geometries are defined by the polar angle $\phi_H$ of the external magnetic field $\mathbf{H}$.
Since the symmetry of the magnetoacoustic driving field $\bf h$ essentially determines the magnitude of the magnetoacoustic interaction, we will now discuss the orientation dependence of $|\mu_0 \tilde{\mathbf{h}} (\theta_0)|$ for the Rayleigh wave strain components $\varepsilon_{xx}$,  $\varepsilon_{zz}$ and  $\varepsilon_{xz}$ separately, setting all other strain components equal to zero~\cite{Dreher.2012}. 
In Fig.~\ref{fig:4} we show a polar plot of the normalized magnitude of the driving field $|\mu_0 \tilde{\mathbf{h}} (\theta_0)|$, using $2 b_{1,2} \tilde{a}_{kl} = \SI{1}{T}$ and assuming no in-plane anisotropy ($H_\text{ani}=0, \phi_0 = \phi_H$). 
First, it is interesting, that magnetoelastic excitation of SWs in the FV-geometry ($\theta_0=\SI{0}{\degree}$) can be solely mediated by the driving fields of the shear component $\varepsilon_{xz}$.
%
%
Second, finite element method (FEM) eigenmode simulations reveal~\cite{Comsol.}, that the strain component $\varepsilon_{zz}$ is phase shifted by $\pi$ with respect to $\varepsilon_{xx}$. Thus, the magnetoacoustic driving fields of $\varepsilon_{xx}$ and $\varepsilon_{zz}$ show a constructive superposition.
Third, the SAW-SW helicity mismatch effect arises because of a $\pm \pi/2$ phase shift of $\varepsilon_{xz}$ with respect to $\varepsilon_{xx}$~\cite{Lewis.1972,Dreher.2012,Sasaki.2017,Tateno.2020,A.HernandezMinguez.2020,Ku.2020,Ku.2021}. Under an inversion of the SAW propagation direction ($k \rightarrow -k$, or $k_{S21} \rightarrow k_{S12}$), the phase shift changes its sign ($\pi/2 \rightarrow -\pi/2$). For measurements in the in-plane geometry, the SAW-SW helicity mismatch effect is attributed to a superposition of driving fields caused by $\varepsilon_{xx}$ and $\varepsilon_{xz}$.
This is in contrast to the oop90-geometry ($\phi_0=\SI{90}{\degree}$), where the SAW-SW helicity mismatch effect is mediated by the strain components $\varepsilon_{zz}$ and $\varepsilon_{xz}$.

The magnetoacoustic driving field causes the excitation of SWs in the magnetic film. Thus, the power of the traveling SAW is exponentially decaying while propagating through the magnetic film with length $l_f$ and thickness $d$. With respect to the initial power $P_0$, the absorbed power of the SAW is
\begin{eqnarray}
&P_\text{abs} = P_0
\left(
1 - \text{exp}
\left\{
- C ~\operatorname{Im}
\left[
 (\mathbf{\tilde{h}})^*
 \bar{\chi}
 \mathbf{\tilde{h}}
\right]
\right\}
\right)
\nonumber \\
& \text{with}~ C = \frac{1}{2} \mu_0 l_f d \left( \frac{k^2}{R} \right).
\label{eq:PabsDGLsol}
\end{eqnarray}
The magnetic susceptibility tensor $\bar{\chi}$ describes the magnetic response to small time-varying magnetoacoustic fields and is calculated as described by Dreher et al.~\cite{Dreher.2012} for arbitrary equilibrium magnetization directions $(\theta_0, \phi_0)$. 
Besides the external magnetic field, exchange coupling, and uniaxial in-plane anisotropy, we take additionally the dipolar fields for SWs with $k \neq 0$ into account, which are given in Eq.~\eqref{eq:dipolar fields} in the Appendix~\ref{appendix:dipolar fields}.

Finally, to directly simulate the experimentally determined relative change of the SAW transmission $\Delta S_{ij}$ on the logarithmic scale, we use
\begin{equation}
\Delta S_{ij}
= 10 \lg \left( \frac{P_0 - P_\text{abs}}{P_0} \right) 
\ \text{with} \  
ij = 
\left\{
\begin{array}{ll}
21, & \text{for}~k \geq 0 \\
12, & \text{for}~k<0 \\
\end{array}
\right.
\label{eq:5:S21Final}
\end{equation}
for SAWs propagating parallel ($k \geq 0$) and antiparallel ($k<0$) to the $x$-axis.

\subsection{Spin wave dispersion}

Resonant SAW-SW excitation is possible if the dispersion relations of SAW and SW intersect in the uncoupled state.
The SW dispersion is obtained by setting $\text{det}\left( \bar{\chi}^{-1} \right) = 0$ and taking the real part of the solution for small SW damping constants $\alpha$. If we neglect the uniaxial in-plane anisotropy ($H_\text{ani}=0$, $\phi_0 = \phi_H$) we obtain~\cite{CortesOrtuno.2013}
\begin{equation}
\label{eq:FWSW dispersion}
f=
    \frac{\gamma \mu_0}{2 \pi}
    \sqrt{
    H_{11} H_{22} - H_{12}^2
    }
\end{equation}
with
\begin{eqnarray}
\label{eq:3}
H_{11}=&&
    H \cos{(\theta_0-\theta_H)}
    +D k^2
    -M_s \cos{(2 \theta_0)} \nonumber\\&&
    +M_s ( 1-G_0 )  \left( \cos{(2 \theta_0)} - \sin^2{\phi_0} \cos^2{\theta_0} \right)
\nonumber\\
H_{22}=&&
    H \cos{(\theta_0-\theta_H)}
    +D k^2
    -M_s \cos^2{\theta_0} \nonumber\\&&
    +M_s ( 1-G_0 ) \sin^2{\phi_0}
\nonumber\\
H_{12}=&&
    M_s ( 1-G_0 ) \sin{\phi_0} \cos{\phi_0} \cos{\theta_0}.
\end{eqnarray}
Here, $\gamma$ is the gyromagnetic ratio, $G_0 = \frac{1-\text{e}^{-|k|d}}{|k|d}$ and $D=\frac{2 A}{\mu_0 M_s}$ with the magnetic exchange constant $A$.

\begin{figure}
\includegraphics[width = 0.48\textwidth]{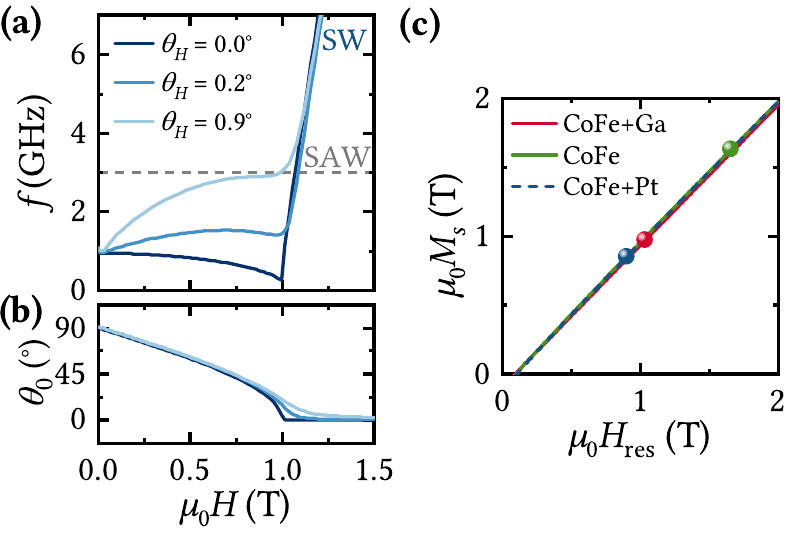}
\caption{
(a) The SW resonance frequency $f$ is calculated with Eq.~\eqref{eq:FWSW dispersion} for the oop0-geometry as a function of the external magnetic field magnitude $\mu_0 H$ and azimuthal angle $\theta_H$.
The corresponding azimuthal angle $\theta_0$ of the equilibrium magnetization orientation is shown in (b). For the simulation, we use $\phi_0=\SI{0}{\degree}$, $k=\SI{5.9}{\micro m^{-1}}$, $\mu_0 M_s = \SI{1}{T}$, and zero in-plane anisotropy. The remaining parameters are taken from the CoFe+Ga thin film in Table~\ref{tab:table2}.
(c) 
The saturation magnetizations $M_s$ of the three different magnetic thin films (colored dots) are calculated from the experimentally determined resonance field $\mu_0 H_\text{res}$ of the FVSW in Fig.~\ref{fig:8}. 
The general dependence $\mu_0 M_s (\mu_0 H_\text{res})$ is shown by the lines for the different magnetic films.
\label{fig:5}
}
\end{figure}

We exemplarily calculated the SW resonance frequency $f$ in Fig.~\ref{fig:5}(a) for the oop0-geometry as a function of the external magnetic field magnitude $\mu_0 H$.
The corresponding azimuthal angle $\theta_0$ of the equilibrium magnetization orientation is shown in Fig.~\ref{fig:5}(b). For the simulation, we use besides $\phi_0=\SI{0}{\degree}$, $k=\SI{5.9}{\micro m^{-1}}$, $\mu_0 M_s = \SI{1}{T}$ and $H_\text{ani}=0$ the parameters of the CoFe+Ga thin film in Table~\ref{tab:table2}.
Additionally, the resonance frequency $f=\SI{3}{GHz}$ of a SAW with $k=\SI{5.9}{\micro m^{-1}}$ is depicted by the dashed line in Fig.~\ref{fig:5}(a).
The dispersion $f(\mu_0 H)$ changes strongly with the azimuthal angle $\theta_H$ of the applied external magnetic field. 
For the FVSW-geometry $\theta_H =\SI{0}{\degree}$, the magnetic thin film is saturated ($\theta_0 =\SI{0}{\degree}$) when the magnetic field overcomes the magnetic shape anisotropy $\mu_0 H > \mu_0 M_s$ and resonant SAW-SW interaction is only possible at $\mu_0 H = \SI{1.06}{T}$. 
In contrast, for $\theta_H =\SI{0.9}{\degree}$, we expect magnetoacoustic interaction in a wide range $\mu_0 H \approx \num{0.7}, ..., \SI{1.0}{T}$, where the dispersions of SAW and SW intersect. For this geometry and $\mu_0 H \leq \SI{1.5}{T}$, the magnetic film is not fully saturated ($\theta_0 \neq \SI{0.9}{\degree}$). 

\section{Experimental Setup}

\begin{figure}
\includegraphics[width = 0.48\textwidth]{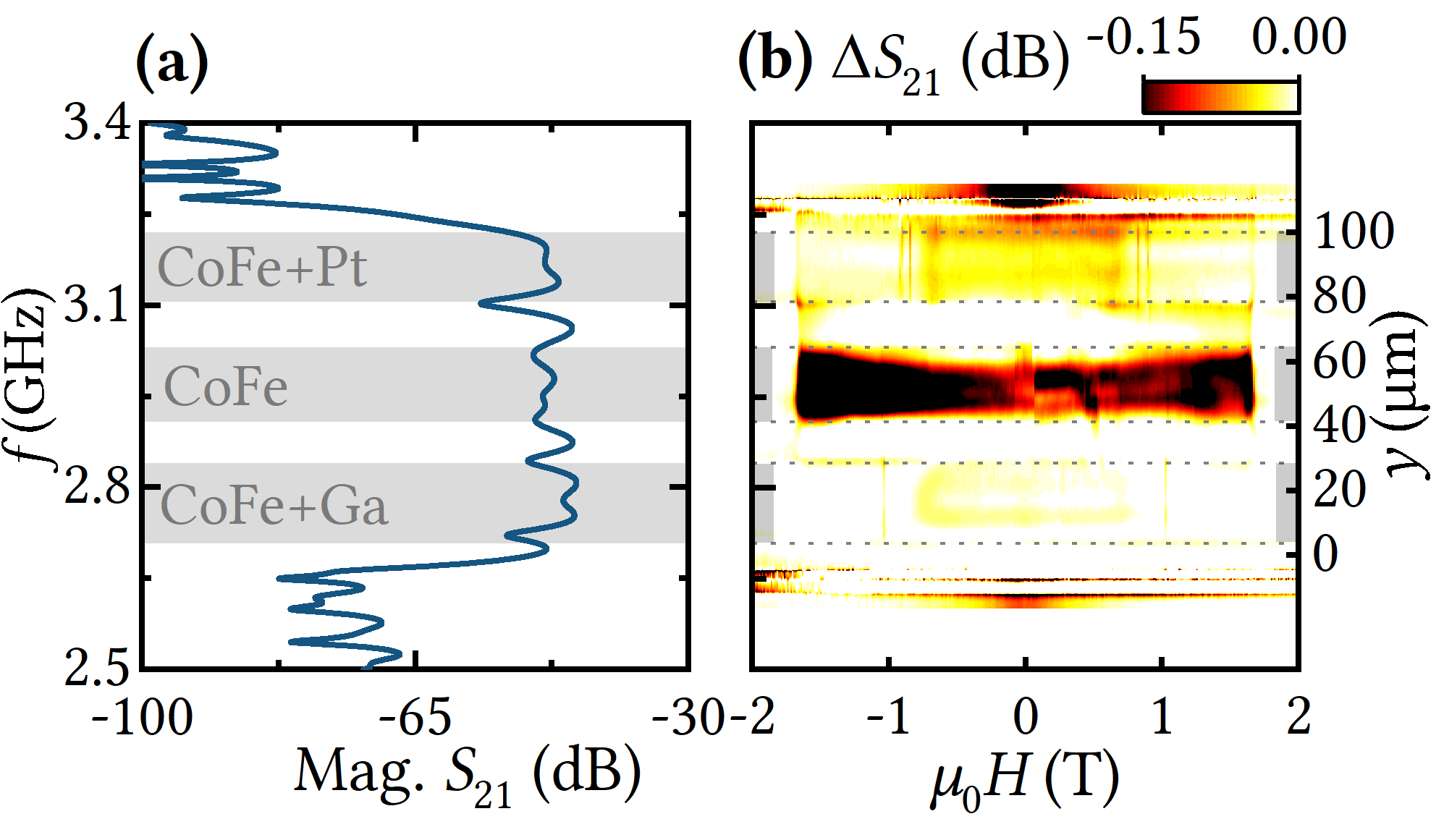}
\caption{
(a) The transmission characteristics of the fabricated device shows the expected wide band behavior.
(b) Within this transmission band, the magnetoacoustic transmission $\Delta S_{21}(\mu_0 H)$ differs for the three different frequency sub-bands, that correspond to the three different magnetic films.
\label{fig:6}
}
\end{figure}

In contrast to previous magneotoacoustic studies performed with conventional IDTs~\cite{Gowtham.2015,A.HernandezMinguez.2020,Xu.2020,Ku.2020,Ku.2021,Ku.2021b,Duquesne.2019,Thevenard.2014}, here we use "tapered" or "slanted" interdigital transducers (TIDTs) \cite{vandenHeuvel.1972, yatsuda1997design, solie1998tapered, Streibl.1999} to characterize SAW-SW interaction in three different magnetic thin micro-stripes in one run. 
Although the fingers of the TIDT are slanted, the SAW propagates dominantly parallel to the $x$-axis in Fig.~\ref{fig:1} because of the strong beam steering effect of the Y-cut Z-propagation LiNbO$_3$ substrate~\cite{vandenHeuvel.1972,Morgan.2007}.
The linear change of the periodicity $p(y)$ along the transducer aperture $W$ results in a spatial dependence of the SAW resonance frequency $f(y)=c_\text{SAW} / p(y)$~\cite{vandenHeuvel.1972}. Thus, a TIDT has a wide transmission band and can be thought of to consist out of multiple conventional IDTs that are connected electrically in parallel~\cite{solie1998tapered}. 
In good approximation, the frequency bandwidth of a conventional IDT is given by $\Delta f_\text{IDT} = 0.9 f_0 / N$ and is constant for higher harmonic resonance frequencies.
From the bandwidth $\Delta f_\text{TIDT}$ of the TIDT the width of the acoustic beam $w$ at constant frequency can be estimated~\cite{Streibl.1999} with
\begin{equation}
    \label{eq:TIDT w}
    w = W \frac{\Delta f_\text{IDT}}{\Delta f_\text{TIDT}}.
\end{equation}
The TIDTs are fabricated out of Ti(5)/Al(70) (all thicknesses are given in units of nm), have an aperture of $W=\SI{100}{\micro m}$, the number of finger-pairs is $N=22$ and the periodicity $p(y)$ changes from \SI{3.08}{\micro m} to \SI{3.72}{\micro m}.
As shown in Fig.~\ref{fig:6}(a), we operate the TIDT at the third harmonic resonance, which corresponds to a transmission band and SAW wave length in the ranges of $\SI{2.69}{GHz} < f < \SI{3.22}{GHz}$ and $\SI{1.06}{\micro m} < \lambda < \SI{1.27}{\micro m}$.
According to Eq.~\eqref{eq:TIDT w}, we expect for the width of the acoustic beam at constant frequency $w = \SI{100}{\micro m} (\SI{41}{MHz} / \SI{530}{MHz}) \approx \SI{7.7}{\micro m}$.
Moreover, Streibel et al. argue that internal acoustic reflections in the single electrode structure used additionally lowers $w$ by about a factor of four~\cite{Streibl.1999}.
%
Since $\lambda$ is in the range of $w$, diffraction effects can be expected. These beam spreading losses are partly compensated by the beam steering effect and the frequency selectivity of the receiving transducer, which filters out the diffracted portions of the SAW~\cite{Streibl.1999}.

The three different magnetic micro-stripes in Fig.~\ref{fig:1} were deposited by direct-writing techniques between the two \SI{800}{\micro m} distant TIDTs. For details we refer to the appendix~\ref{appendix:FEBID FIBID details}. The compositions of the deposited magnetic films were characterized by energy-dispersive X-ray spectroscopy (EDX). The results are summarized in Table~\ref{tab:table1}. 
More details about the microstructure and magnetic properties of CoFe can be found in Refs.~\cite{Bunyaev.2021, Keller.2018}. For the microstructure of mixed CoFe-Pt deposits we refer to Ref.~\cite{Porrati.2012} in which results of a detailed investigation of the microstructural and magnetic properties of fully analogous Co-Pt deposits are presented. 
We determined the thicknesses $d$ and the root mean square roughness of the samples CoFe+Ga($24 \pm 2$), CoFe($72 \pm 2$), and CoFe+Pt($70 \pm 2$) by atomic force microscopy (AFM). The length and widths of all micro-stripes are identical with $l_f=\SI{40}{\micro m}$ and $w_f = \SI{20}{\micro m}$, except $w_f^\text{CoFe+Ga} = \SI{26}{\micro m}$.

\begin{table}
\caption{
\label{tab:table1}
Compositional EDX analysis of test samples with size $\SI{1.5}{\micro m} \times \SI{1.5}{\micro m}$.
The electron beam voltage was \SI{5}{keV} for FEBID samples and \SI{3}{keV} for FIBID sample.
}
\begin{ruledtabular}
\begin{tabular}{ccccccc}
Sample & C & O & Fe & Co & Ga & Pt \\
\colrule
CoFe+Pt & 61.8 & 6.5 &  4.2 & 20.1 &  & 7.4 \\
CoFe    & 26.2 & 6.9 & 12.4 & 54.5 &  &  \\
CoFe+Ga & 16.9 & 16.5 &  7.7 & 37.5 & 21.4 &  \\
\end{tabular}
\end{ruledtabular}
\end{table}

The SAW transmission of our delay line device was characterized by a vector network analyzer. Based on the low propagation velocity of the SAW, a time-domain gating technique was employed to exclude spurious signals~\cite{Hiebel.2011}, in particular electromagnetic crosstalk.
We use the relative change of the background-corrected SAW transmission signal as
\begin{equation}
    \Delta S_{ij} (\mu_0 H) = S_{ij} (\mu_0 H) - S_{ij} (\SI{2}{T})
\end{equation}
to characterize SAW-SW coupling. Here $\Delta S_{ij}$ is the magnitude of the complex transmission signal with $ij \in \{21,12\}$. In all measurements, the magnetic field is swept from \SI{-2}{T} to \SI{+2}{T}.

\section{Discussion}


\subsection{Experimental results}

In Fig.~\ref{fig:6}(b), we show the magnetoacoustic transmission $\Delta S_{21}$ as a function of external magnetic field magnitude and frequency for the FVSW-geometry ($\theta_H \approx \SI{0}{\degree}$).
Within the wide transmission band of the TIDT, the magnetoacoustic transmission $\Delta S_{21}(\mu_0 H)$ clearly differs for the three different frequency sub-bands, each of which spatially addresses one of the three different magnetic micro-stripes. 
Both, the maximum change of the transmission with $\text{Max}(\Delta S_{21}^{\text{CoFe}}) > \text{Max}(\Delta S_{21}^{\text{CoFe+Pt}}) > \text{Max}(\Delta S_{21}^{\text{CoFe+Ga}})$ and the resonance fields are different for the three films.
The small signals $\Delta S_{21} \neq 0$ at frequencies corresponding to the gaps between the magnetic structures are attributed to diffraction effects. The apparent signal $\Delta S_{21}$ at the edges of the transmission band is attributed to measurement noise.
%
From Fig.~\ref{fig:6}(b) we identify the frequencies which correspond to the centers of the three magnetic films CoFe+Ga, CoFe, and CoFe+Pt as \SI{2.78}{GHz}, \SI{2.96}{GHz}, and \SI{3.17}{GHz}, respectively.
Further analysis is performed at these fixed frequencies.

\begin{figure*}
\includegraphics[width = 0.95\textwidth]{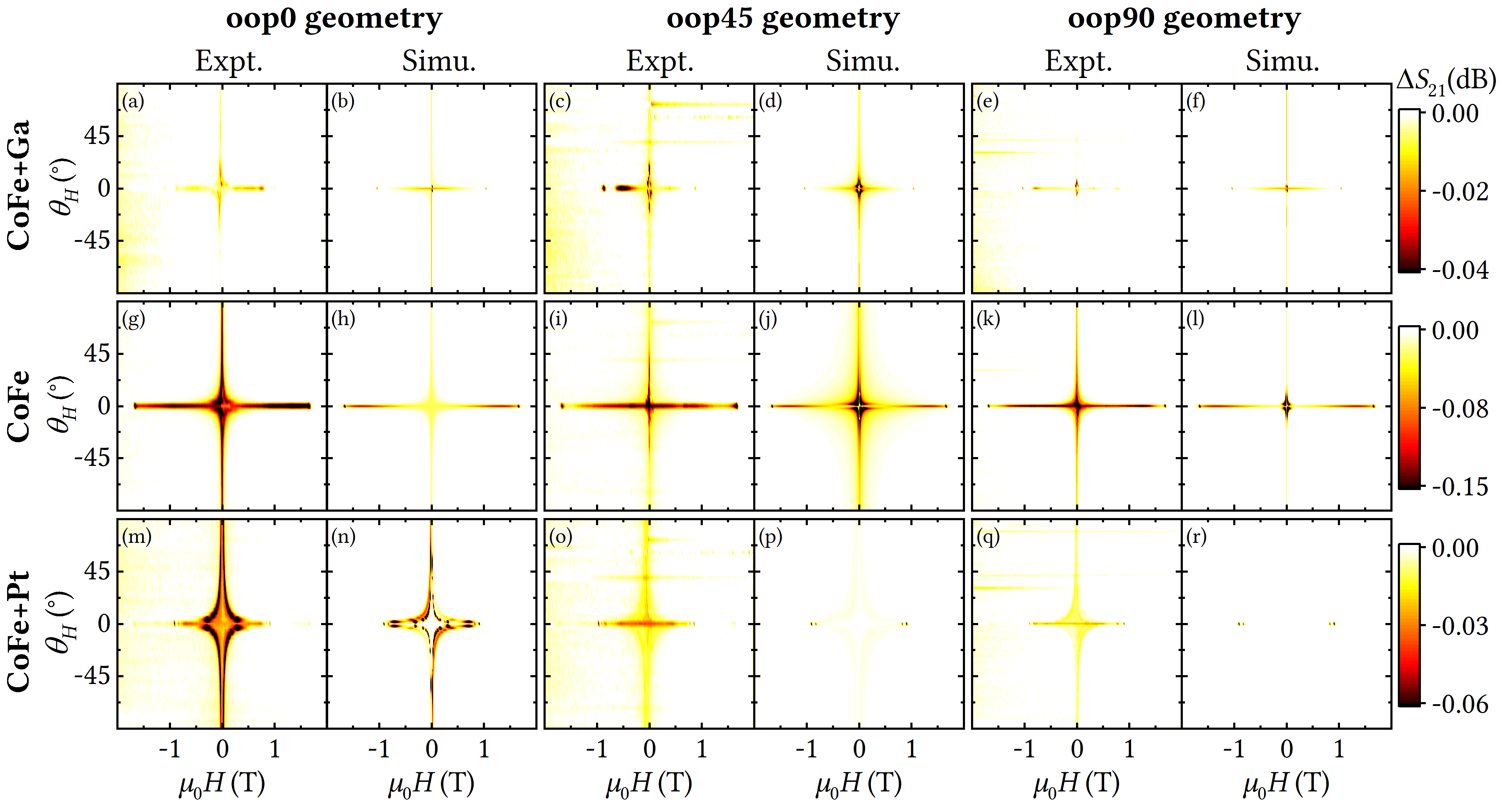}
\caption{
The magnetoacoustic transmission $\Delta S_{21}(\mu_0 H,\theta_H)$ of the magnetic micro-stripes CoFe+Ga (\SI{2.78}{GHz}), CoFe (\SI{2.96}{GHz}), and CoFe+Pt (\SI{3.17}{GHz}) is shown in the oop0-, oop45-, and oop90-geometry (see Fig.~\ref{fig:3}).
Resonances are observed for $\theta_H=\SI{0}{\degree}$, which are studied in more detail in Fig.~\ref{fig:7}.
Simulation and experiment show good qualitative agreement.
\label{fig:7}
}
\end{figure*}

\begin{figure*}
\includegraphics[width = 0.95\textwidth]{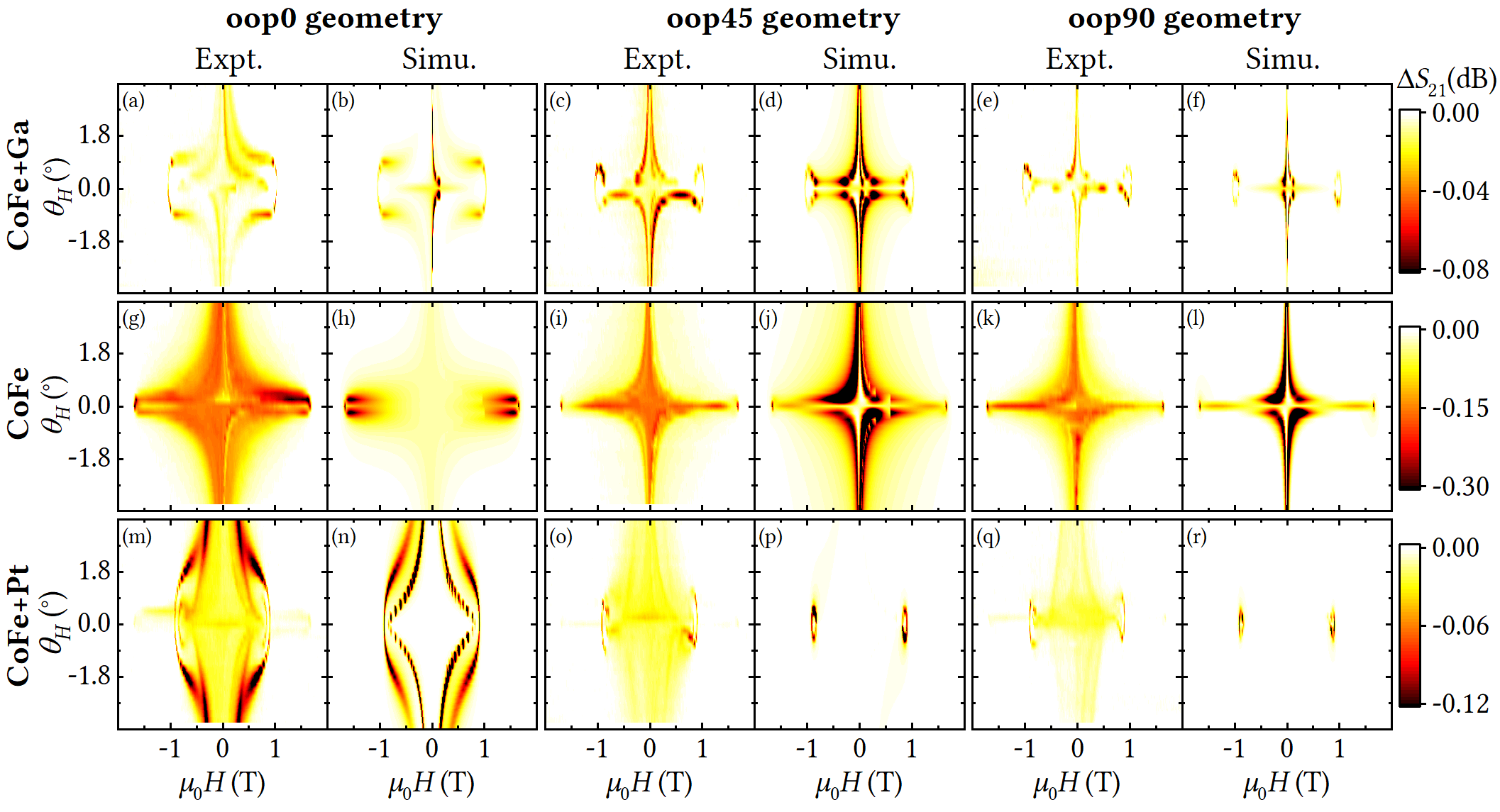}
\caption{
The magnetoacoustic transmission $\Delta S_{21}(\mu_0 H,\theta_H)$ of the magnetic micro-stripes CoFe+Ga (\SI{2.78}{GHz}), CoFe (\SI{2.96}{GHz}), and CoFe+Pt (\SI{3.17}{GHz}) is shown in the oop0-, oop45-, and oop90-geometry (see Fig.~\ref{fig:3}) for almost out-of-plane oriented external magnetic field ($\theta_H=\SI{-3,6}{\degree}, \ldots, \SI{+3,6}{\degree}$).
Simulation and experiment show good qualitative agreement.
\label{fig:8}
}
\end{figure*}

In Fig.~\ref{fig:7}, we show the magnetoacoustic transmission $\Delta S_{21}(\mu_0 H,\theta_H)$ of all three films in the oop0-, oop45-, and oop90-geometry (see Fig.~\ref{fig:3}) as a function of external magnetic field magnitude $\mu_0 H$ and orientation $\theta_H$ in a range of $\SI{-90}{\degree} \leq \theta_H \leq \SI{+90}{\degree}$ with an increment of $\Delta \theta_H = \SI{3.6}{\degree}$.
For almost all geometries, the magnetoacoustic response $\Delta S_{21}(\mu_0 H,\theta_H)$ has a star shape symmetry, which was already observed by Dreher et al.\ for Ni(50) thin films~\cite{Dreher.2012}. This symmetry results from magnetic shape anisotropy.
The sharp resonances in Fig.~\ref{fig:7} around $\theta_H=\SI{0}{\degree}$ are studied in Fig.~\ref{fig:8} in the range of $\SI{-3,6}{\degree} \leq \theta_H \leq \SI{+3,6}{\degree}$ with $\Delta \theta_H = \SI{0.225}{\degree}$ in more detail. For all three magnetic micro-stripes SWs can be magnetoacoustically excited in the FVSW-geometry ($\theta_H=\SI{0}{\degree}$) and the resonance fields $\mu_0 H_\text{res}(\theta_H=\SI{0}{\degree})$ differ.
Additionally, the symmetry of the magnetoacoustic resonances $\mu_0 H_\text{res}(\theta_H)$ changes for the geometries oop0, oop45, and oop90 and the different magnetic micro-stripes. 
In general, the resonance fields $|\mu_0 H_\text{res}|$ decrease if $|\phi_H|$ is increased from \SI{0}{\degree} to \SI{90}{\degree} (oop0 to oop90).
Moreover, the line symmetry with respect to $\theta_H = \SI{0}{\degree}$ is broken, in particular for the oop45-, and oop90-geometry.

\subsection{Simulation and Interpretation}

To simulate the experimental results in Figs.~\ref{fig:7} and \ref{fig:8} with Eq.~\eqref{eq:5:S21Final}, we first have to determine the saturation magnetizations $M_s$ of the different magnetic thin films.
For this purpose, we compute Eq.~\eqref{eq:FWSW dispersion} for the FVSW geometry ($\theta_H = \SI{0}{\degree}, \theta_0 = \SI{0}{\degree}$). The relation $M_s(H \equiv H_\text{res})$ is shown in Fig.~\ref{fig:5}(c) for all three magnetic films. Thereby, frequency $f$ and wave vector $k$ of the SW are determined by the SAW and we assume $c_\text{SAW}= \SI{3200}{m/s}$~\cite{Note1}, $g=\num{2.18}$~\cite{Bunyaev.2021} and $D=\SI{24.7E-12}{A m}$~\cite{Bunyaev.2021}.
Since the in-plane anisotropy $H_\text{ani}$ is expected to be small compared to the shape anisotropy, the impact on the resonance in the FVSW geometry is small, and we use $H_\text{ani} = 0$.
Under these assumptions, the relations $M_s(H_\text{res})$ are almost identical for the three magnetic films.
Together with the experimentally determined $\mu_0 H_\text{res}(\theta_H=\SI{0}{\degree})$ in Fig.~\ref{fig:8}, the saturation magnetizations of CoFe+Ga, CoFe, and CoFe+Pt are determined to be \SI{772}{kA/m}, \SI{1296}{kA/m}, and \SI{677}{kA/m}.
%


For the simulations in Figs.~\ref{fig:7} and \ref{fig:8}, we use the parameters summarized in Table~\ref{tab:table2}.
The complex amplitudes of the normalized strain $\tilde{a}_{kl}=\varepsilon_{kl,0}/{|k||u_{z,0}|}$ are estimated from a COMSOL~\cite{Comsol.} finite element method (FEM) simulation.
Since we do not know the elastic constants and density of the magnetic micro-stripes, we assume a pure LiNbO$_3$ substrate with a perfectly conducting overlayer of zero thickness.
Thus, the real values of $\tilde{a}_{kl}$ might deviate from the assumed ones~\cite{Ku.2020}. Furthermore, the normalized strain of the simulation was averaged over the thickness $0 \leq z \leq -d$. 
The values for the SW effective damping $\alpha$, magnetoelastic coupling for polycrystalline films~\cite{Dreher.2012} $b_1=b_2$ and small phenomenological uniaxial in-plane anisotropy ($H_\text{ani}$, $\phi_\text{ani}$) were adjusted to obtain a good agreement between experiment and simulation. 
Thereby, $\alpha$ includes Gilbert damping and inhomogeneous line broadening~\cite{Ku.2020}. The phenomenological uniaxial in-plane anisotropy could be caused by substrate clamping effects or the patterning strategy of the FEBID / FIBID direct-write process. Note that the values of all these parameters listed in Table~\ref{tab:table2} are very reasonable.

\begin{table}
\caption{
\label{tab:table2}
Parameters to simulate the magnetoacoustic transmission $\Delta S_{21}$ ($k>0$) of the Rayleigh-type SAW in Figs.~\ref{fig:7}-\ref{fig:9}. For the simulation of $\Delta S_{12}$ ($k<0$), the sign of the normalized strain $\tilde{a}_{xz}$ is inverted. For all micro-stripes, we assume $g=\num{2.18}$~\cite{Bunyaev.2021} and $D=\SI{24.7E-12}{A m}$~\cite{Bunyaev.2021}.
}
\begin{ruledtabular}
\begin{tabular}{cccc}
  & CoFe+Ga & CoFe & CoFe+Pt    \\
\colrule
$d$ \, \si{(nm)} & 24 & 72 & 70 \\ 
$f$\,\si{(GHz)} & 2.78 & 2.96 & 3.17 \\ 
$M_s$\,\si{(kA/m)} & 772 & 1296 & 677 \\ 
$\alpha$ & 0.04 & 0.1 & 0.05 \\ 
$\phi_\text{ani}$\,\si{(\degree)} & -10 & 0 & 88 \\ 
$\mu_0 H_\text{ani}$\,\si{(mT)} & 1 & 5 & 10 \\ 
$\tilde{a}_{xx}$ & 0.49 & 0.40 & 0.40 \\ 
$\tilde{a}_{zz}$ & -0.15 & -0.10 & -0.10 \\
$\tilde{a}_{xz}$ & 0.13i & 0.17i & 0.17i \\ 
$|b_1|$\,\si{(T)} & 4 & 15 & 6 \\ 
\end{tabular}
\end{ruledtabular}
\end{table}


For all three magnetic micro-stripes, the qualitative agreement between simulation and experiment in Figs.~\ref{fig:7} and \ref{fig:8} is good.
For magnetoelastic interaction, SWs can be excited in the FVSW-geometry ($\theta_H=\SI{0}{\degree}$) solely due to the vertical shear strain $\varepsilon_{xz}$ which causes a nonzero magnetoacoustic driving field, as discussed in Fig.~\ref{fig:4}.
According to Eq.~\eqref{eq:drivingfields2} the driving field mediated by $\varepsilon_{xx,zz}$ contributes for $\theta_H \neq \SI{0}{\degree}$. 
In Fig.~\ref{fig:8}, the intensity of the resonances for $\theta_H \neq \SI{0}{\degree}$ is therefore more pronounced than for $\theta_H = \SI{0}{\degree}$.
Because the driving fields, which are mediated by the strain $\varepsilon_{xx}$ and $\varepsilon_{zz}$, are in phase, SW excitation in one of the out-of-plane geometries can be even more efficient than in the in-plane geometry.
%
The magnetoacoustic resonance fields of the three magnetic micro-stripes mainly differ, due to differences in $M_s$ and $d$, which strongly affect the corresponding dipolar fields of a SW.
As expected from the SW dispersion in Fig.~\ref{fig:5}(a), we observe for the CoFe+Ga film in Fig.~\ref{fig:8}(a,b) for $\theta_H=0$ a resonance at $\mu_0 H = \SI{1.06}{T}$ with a narrow linewidth and for $\theta_H=\SI{0.9}{\degree}$ a wide resonance between $\mu_0 H \approx \num{0.7}, ..., \SI{1.0}{T}$.
The symmetry of the magnetoacoustic resonances $\mu_0 H_\text{res}(\theta_H)$ changes with the geometries oop0, oop45 and oop90 since the magnetic dipolar fields of the SW dispersion Eq.~\eqref{eq:FWSW dispersion} depend on $\phi_0$.
For CoFe+Pt, two resonances are observed in the oop00-geometry, whereas in the oop45- and oop90-geometry confined oval-shaped resonances show up. This behavior can be modeled by assuming an uniaxial in-plane anisotropy with $\phi_\text{ani} \approx \SI{90}{\degree}$.
In the oop00-geometry, the resonance with the lower resonant fields can be attributed to the switching of the in-plane direction of the equilibrium magnetization direction. In the oop45- and oop90-geometries, the resonance frequencies of the SWs are higher than the excitation frequency of the SAW for $|\theta_H| > \SI{0.7}{\degree}$.
Thus, the magnetoacoustic response $\Delta S_{21}$ is low for $|\theta_H| > \SI{0.7}{\degree}$ in Figs.~\ref{fig:8}(o)-(r).

We attribute discrepancies between experiment and simulation to the following effects: The phenomenological model solely considers an in-plane uniaxial anisotropy. Additional in- and out-of-plane anisotropies would result in a shift of the resonance fields. Furthermore, the strain is estimated by a simplified FEM simulation and assumed to be homogeneous along the thickness of the micro-stripe. Moreover, we neglect magneto-rotation coupling~\cite{Maekawa.1976, Xu.2020, Ku.2020}, spin-rotation coupling~\cite{Matsuo.2011, Matsuo.2013, Kobayashi.2017} and gyromagnetic coupling~\cite{Kurimune.2020}. These assumptions have an impact on the intensity and symmetry of the resonances.
Finally, low-intensity spurious signals are caused by SAW diffraction effects which are, for instance, observed in Fig.~\ref{fig:8}(m,o,q) for $|\mu_0 H| > \SI{1}{T}$.

\subsection{Nonreciprocal behavior}

\begin{figure*}
\includegraphics[width = 0.95\textwidth]{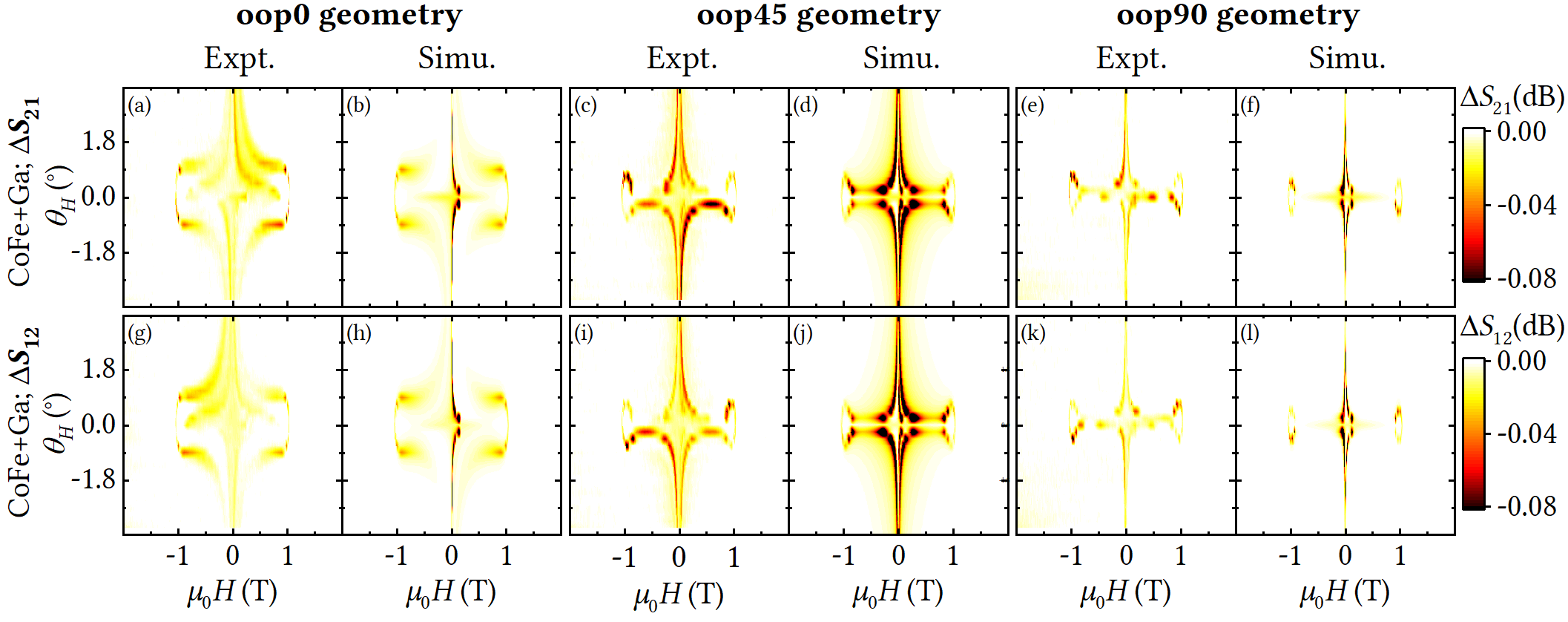}
\caption{
Nonreciprocal magnetoacoustic waves are characterized by different transmission amplitudes $\Delta S_{21}$ and $\Delta S_{12}$ for oppositely propagating SAWs with wave vectors $k_\text{S21}$ and $k_\text{S12}$.
The nonreciprocal transmission is exemplarily shown for the magnetic micro-stripes CoFe+Ga (\SI{2.78}{GHz}) in the oop0-, oop45-, and oop90-geometry for almost out-of-plane oriented external magnetic field ($\theta_H=\SI{-3,6}{\degree}, \ldots, \SI{+3,6}{\degree}$).
Nonreciprocal behavior can solely be observed in the oop45- and oop90-geometry, which is nicely reproduced by the simulation.
\label{fig:9}
}
\end{figure*}

The nonreciprocal behavior of the magnetoacoustic wave in the oop0, oop45, and oop90-geometries is exemplarily shown for CoFe+Ga in Fig.~\ref{fig:9}.
If the magnetoacoustic wave propagates in inverted directions $k_{S21}$ and $k_{S12}$ ($k$ and $-k$) the magnetoacoustic transmission $\Delta S_{21}(\mu_0 H, \theta_H)$ and $\Delta S_{12}(\mu_0 H, \theta_H)$ differs for the oop45- and oop90-geometry. The qualitative agreement between experiment and simulation is also good with respect to the nonreciprocity.
The SAW-SW helicity mismatch effect, discussed in the theory section, causes $\Delta S_{21}(\mu_0 H, \theta_H) \neq \Delta S_{12}(\mu_0 H, \theta_H)$ in Fig.~\ref{fig:9} and the broken line symmetry with respect to $\theta_H = \SI{0}{\degree}$ in Figs.~\ref{fig:8} and \ref{fig:9}.
So far, nonreciprocal magnetoacoustic transmission was only observed in studies where the external magnetic field was aligned in the plane of the magnetic film ($\theta_H=\SI{90}{\degree}$)~\cite{Lewis.1972,Dreher.2012,Sasaki.2017,Tateno.2020,A.HernandezMinguez.2020,Ku.2020,Ku.2021}.
The magnetoacoustic driving field in Eq.~\eqref{eq:drivingfields2} is linearly polarized along the $1$-axis for $\phi_0=0$. Thus, no nonreciprocity due to the SAW-SW helicity mismatch effect is observed in the oop0-geometry.
In contrast, the driving field has a helicity in the oop45- and oop90-geometry. Since this helicity is inverted under inversion of the propagation direction of the SAW ($\varepsilon_{xz,0} \rightarrow -\varepsilon_{xz,0}$), nonreciprocal behavior shows up in the oop45- and oop90-geometry.
In comparison to the experimental results, the simulation slightly underestimates the nonreciprocity. This is mainly attributed to magneto-rotation coupling~\cite{Maekawa.1976, Xu.2020, Ku.2020}, which can be modeled by a modulated effective coupling constant $b_{2,\text{eff}}$ and can result in an enhancement of the SAW-SW helicity mismatch effect~\cite{Xu.2020,Ku.2020}.

\section{Conclusions}

In conclusion, we have demonstrated magnetoacoustic excitation and characterization of SWs with micron-scale spatial resolution using TIDTs.
The magnetoacoustic response at different frequencies, which lay within the wide transmission band of the TIDT, can be assigned to the spatially separated CoFe+Ga, CoFe, and CoFe+Pt magnetic micro-stripes.
%
SAW-SW interaction with micron-scale spatial resolution can be interesting for future applications in magnonics and the realization of new types of microwave devices such as magnetoacoustic sensors~\cite{Chiriac.2001, A.Kittmann.2018, Muller.2022} or microwave acoustic isolators~\cite{R.Verba.2019, Shah.2020, Ku.2021b, Matsumoto.2022}.
For instance, giant nonreciprocal SAW transmission was observed in magnetic bilayers and proposed to build reconfigurable acoustic isolators~\cite{R.Verba.2019, Shah.2020, Ku.2021b, Matsumoto.2022}. In combination with TIDTs, acoustic isolators, which show in adjacent frequency bands different nonreciprocal behavior could be realized. 
Furthermore, if two orthogonal delay lines are combined in a cross-shaped structure, resolution of magnetoacoustic interaction of different magnetic micro-structures in two dimensions can potentially be achieved~\cite{Streibl.1999, Paschke.2017}.

In addition, we extended the theoretical model of magnetoacoustic wave transmission~\cite{Dreher.2012, Ku.2020} in terms of SWs with nonzero wave vector and arbitrary out-of-plane orientation of the static magnetization direction. This phenomenological model describes the experimental results for CoFe+Ga, CoFe, and CoFe+Pt magnetic micro-stripes in different geometries of the external magnetic field - including the FVSW-geometry - in a good qualitative way.
We find that FVSWs can be magnetoelastically excited by Rayleigh-type SAWs due to the shear strain component $\varepsilon_{xz}$. Also magneto-rotation coupling~\cite{Maekawa.1976, Xu.2020, Ku.2020}, spin-rotation coupling~\cite{Matsuo.2011, Matsuo.2013, Kobayashi.2017} or gyromagnetic coupling~\cite{Kurimune.2020} may contribute to the excitation of FVSWs. Since the SAW-SW helicity mismatch effect, which is related to $\varepsilon_{xz}$ and the effective coupling constant $b_{2,\text{eff}}$, is low in Ni thin films~\cite{Weiler.2009,Dreher.2012,Sasaki.2017, Gowtham.2015,Labanowski.2016}, we expect a low excitation efficiency for FVSWs in Ni.
In contrast to the previously discussed in-plane geometry, the strain component $\varepsilon_{zz}$ of Rayleigh-type waves plays an important role in the out-of-plane geometries and can result in enhanced SAW-SW coupling efficiency and SAW-SW helicity mismatch effect.

\begin{acknowledgments}
This work is funded by the Deutsche Forschungsgemeinschaft (DFG, German Research Foundation) – project numbers 391592414 and 492421737.
M.H. acknowledges support by the Deutsche Forschungsgemeinschaft (DFG) through the trans-regional collaborative research center TRR 288 (project A04) and through project No. HU 752/16-1.
\end{acknowledgments}


\appendix

\section{Effective dipolar fields}
\label{appendix:dipolar fields}

The effective dipolar fields in the (1,2,3) coordinate system for arbitrary equilibrium magnetization directions $(\theta_0, \phi_0)$ are taken from Ref.~\cite{CortesOrtuno.2013} by comparing Eq.~23 with the Landau--Lifshitz equation
\begin{eqnarray}
&&
\mathbf{H}_{\text{eff},123}^{\text{dip}}
    =
    M_s
    \begin{pmatrix}
        H_{11}^{\text{dip}} m_1 + H_{12}^{\text{dip}} m_2 \\
        H_{22}^{\text{dip}} m_2 + H_{21}^{\text{dip}} m_1 \\
        -\cos^2{\theta_0}
    \end{pmatrix}
    \nonumber\\
&&
H_{11}^{\text{dip}}
    =
    -\cos^2{\theta_0}
    +\cos{(2 \theta_0)}
    \nonumber\\
    &&
    \hspace{1.2 cm}+( 1-G_0 )  \left(-\cos{(2 \theta_0)} + \sin^2{\phi_0} \cos^2{\theta_0} \right)
    \nonumber\\
&&
H_{22}^{\text{dip}}
    =
    -( 1-G_0 ) \sin^2{\phi_0}
    \nonumber\\
&&H_{12}^{\text{dip}} = H_{21}^{\text{dip}} =  ( 1-G_0 ) \sin{\phi_0} \cos{\phi_0} \cos{\theta_0}.
\label{eq:dipolar fields}
\end{eqnarray}
Here, $m_{1,2}$ are the precession amplitudes of the normalized magnetization $\mathbf{m} = \mathbf{M} / M_s$.

\section{Details about the deposition of the magnetic thin films}
\label{appendix:FEBID FIBID details}

FEBID and FIBID are direct-write lithographic techniques for the fabrication of samples of various dimension, shape and composition~\cite{Huth.2021}. In FEBID/FIBID, the adsorbed molecules of a precursor gas injected in a SEM/FIB chamber dissociate by the interaction with the electron/ion beam forming the sample during the rastering process~\cite{Huth.2018}. In the present work, the samples were fabricated in a dual beam SEM/FIB microscope (FEI, Nova NanoLab 600) equipped with a Schottky electron emitter. FEBID was employed to fabricate the CoFe and CoFe+Pt samples with the following electron beam parameters: \SI{5}{kV} acceleration voltage, \SI{1.6}{nA} beam current, \SI{20}{nm} pitch, and \SI{1}{\micro s} dwell time. The number of passes, i.e., the number of rastering cycles, was 1500. FIBID was used to prepare the CoFe+Ga sample with the following ion beam parameters: \SI{30}{kV} acceleration voltage, \SI{10}{pA} ion beam current, \SI{12}{nm} pitch, \SI{200}{ns} dwell time, and 500 passes. The precursor HFeCo$_3$(CO)$_{12}$ was employed to fabricate the CoFe and the CoFe+Ga samples~\cite{Porrati.2015}, while HFeCo$_3$(CO)$_{12}$ and  (CH$_3$)$_3$CH$_3$C$_5$H$_4$Pt were simultaneously used to grow CoFe+Pt~\cite{Sachser.2021}. Standard FEI gas-injection-systems (GIS) were used to flow the precursor gases in the SEM via capillaries with \SI{0.5}{mm} inner diameter. The distance capillary-substrate surface was about \SI{100}{\micro m} and \SI{1000}{\micro m} for the HFeCo$_3$(CO)$_{12}$ and  (CH$_3$)$_3$CH$_3$C$_5$H$_4$Pt GIS, respectively . The temperature of the precursors were \SI{64}{\celsius} and \SI{44}{\celsius} for  HFeCo$_3$(CO)$_{12}$ and  (CH$_3$)$_3$CH$_3$C$_5$H$_4$Pt, respectively. The  basis pressure of the SEM was \SI{5E7}{mbar}, which rose up to about \SI{6E7}{mbar}, during CoFe and CoFe+Ga deposition, and to about \SI{2E6}{mbar}, during CoFe+Pt deposition. 
\SI{}{}


%

\end{document}